\documentclass[12pt]{article}
\usepackage{amsmath}
\usepackage{graphicx}
\usepackage{natbib}
\usepackage{setspace}
\usepackage{authblk}
\usepackage{hyperref}
\usepackage{geometry}
\usepackage{booktabs}  
\usepackage{multirow}
\usepackage{caption}
\usepackage{amsfonts}
\usepackage{longtable}
\newcommand{\alr}{{\textnormal{alr}}}

\newcommand{\bfX}{\pmb{X}}

\newcommand{\bfbeta}{\pmb{\beta}}
\newcommand{\bftheta}{\pmb{\theta}}
\newcommand{\bfgamma}{\pmb{\gamma}}

\newcommand{\airbnbp}{Airbnb}

\newcommand{\bfy}{\mathbf{y}}
\newcommand{\Dirichlet}{{\textnormal{Dirichlet}}}

\newcommand{\bfmu}{\pmb{\mu}}
\newcommand{\bfeta}{\pmb{\eta}}

\title{Sensitivity Analysis of Priors in the Bayesian Dirichlet Auto-Regressive Moving Average Model}

\author[1,2]{Harrison Katz}
\author[2]{Liz Medina}
\author[3]{Robert E. Weiss}
\affil[1]{Department of Statistics, UCLA}
\affil[2]{Data Science, Forecasting, \airbnbp}
\affil[3]{Department of Biostatistics, UCLA Fielding School of Public Health}

\date{}

\begin{document}

\maketitle

\doublespacing
\begin{abstract}
We examine how prior specification affects the Bayesian Dirichlet Auto-Regressive Moving Average (B-DARMA) model for compositional time series. Through three simulation scenarios—correct specification, overfitting, and underfitting—we compare five priors: informative, horseshoe, laplace, mixture of normals, and hierarchical. Under correct model specification, all priors perform similarly, although the horseshoe and hierarchical priors produce slightly lower bias. When the model overfits, strong shrinkage—particularly from the horseshoe prior—proves advantageous. However, none of the priors can compensate for model misspecification if key VAR/VMA terms are omitted.

We apply B-DARMA to daily S\&P 500 sector trading data, using a large-lag model to demonstrate overparameterization risks. Shrinkage priors effectively mitigate spurious complexity, whereas weakly informative priors inflate errors in volatile sectors. These findings highlight the critical role of carefully selecting priors and managing model complexity in compositional time-series analysis, particularly in high-dimensional settings.
\end{abstract}

\section{Introduction}

Compositional time series, in which observations are vectors of proportions constrained to sum to one, arise in a wide range of applications. For instance, market researchers track evolving shares of competing products \citep{Boonen2019}, ecologists monitor species composition over time, and sociologists or political scientists follow shifting demographic or budgetary profiles \citep{Lipsmeyer2019}. In each case, the data lie within a simplex, making the Dirichlet model a common starting point \citep{Aitchison1986,Greenacre2018}. However, when temporal dependence is also present---for example, today's composition influences tomorrow’s---the Dirichlet framework must be extended to account for dynamics in a way that respects the simplex constraints.

Several approaches have emerged to handle dynamic compositional data, notably by coupling the Dirichlet with Auto-Regressive Moving Average (ARMA)-like structures \citep{zheng2017dirichlet}, using a logistic-normal representation \citep{Casarin2021}, or adopting new families of innovation distributions \citep{Makgai2021}. Others have adapted the model to cope with zeros \citep{Dong2025} or extreme heavy-tailed behaviors. In particular, the \emph{Bayesian Dirichlet Auto-Regressive Moving Average} (B-DARMA) model \citep{KATZ2024} addresses key compositional modeling challenges by introducing Vector Auto-Regression (VAR) and Vector Moving Average (VMA) terms for multivariate compositional data. Under B-DARMA, each day’s, or time point’s, composition is Dirichlet-distributed with parameters that evolve with a VARMA process in the additive log-ratio space, capturing both compositional constraints and serial correlation.

Although B-DARMA provides a flexible foundation, practitioners must still specify priors for potentially high-dimensional parameter spaces, which can be prone to overfitting or omitted-lag bias. The growing literature on Bayesian shrinkage priors offers numerous solutions: from global--local shrinkage frameworks \citep{Griffin2021} to hierarchical approaches \citep{Bitto2019}, along with classic spike-and-slab \citep{Mitchell1988, Follett2019} and Laplace priors \citep{Park2008}. These priors can encourage sparsity and suppress extraneous lags in over-parameterized models, a crucial feature for many real-world compositional applications in which the number of possible lags or covariates exceeds sample size. The horseshoe prior \citep{Carvalho2010, Polson2019} has shown particular promise in forecasting studies where only a minority of parameters matter and many are effectively zero. At the same time, hierarchical shrinkage \citep{Bitto2019} facilitates partial pooling across related coefficient blocks, an appealing property when working with multi-sector or multi-species compositions.

In this paper, we systematically investigate how five different prior families---informative normal, horseshoe, Laplace, spike-and-slab, and hierarchical shrinkage---affect parameter recovery and predictive accuracy in the B-DARMA model. We compare their performance across three main scenarios using simulated data: (i) \emph{correct specification}, where the model order matches the true process, (ii) \emph{overfitting}, where extraneous VAR/VMA terms inflate model dimensionality, and (iii) \emph{underfitting}, where key VAR/VMA terms are missing altogether. Our findings confirm that shrinkage priors---especially horseshoe and hierarchical variants---can successfully rein in overfitting, providing more robust parameter estimates and improved forecasts. Conversely, no amount of shrinkage compensates for omitted terms, highlighting the need for careful model specification. 

To demonstrate practical impact, we also apply B-DARMA to daily S\&P 500 sector trading data, a large-scale compositional time series characterized by multiple seasonalities and long-lag behavior. Consistent with the simulation insights, we find that more aggressive shrinkage priors significantly reduce spurious complexity and improve forecast accuracy, especially for volatile sectors. These outcomes reinforce that while B-DARMA provides a natural scaffolding for compositional dependence, judicious prior selection and meaningful lag choices are pivotal. 

In the remainder of the paper, we first review the B-DARMA model (Section~2) and outline how our five prior families (informative, horseshoe, Laplace, spike-and-slab, and hierarchical) encode distinct shrinkage behaviors. We then present the design and results of three simulation studies (Sections~3-4) before turning to the empirical S\&P 500 application (Section~5). We conclude with recommendations for practitioners modeling compositional time series with complex temporal dynamics and large parameter counts.

\section{Background}

\subsection{Compositional Data and the Dirichlet Distribution}
Compositional data consist of vectors of proportions, each strictly between zero and one and summing to unity \citep{Aitchison1986}. Formally, let
\[
\mathbf{y}_t \;=\; \bigl(y_{t1}, y_{t2}, \ldots, y_{tJ}\bigr)^\prime,\quad t = 1,\dots,T,
\]
where each \(y_{tj}>0\) and \(\sum_{j=1}^J y_{tj} = 1.\) The vector \(\mathbf{y}_t\) resides in the \((J-1)\)-dimensional simplex.

A natural choice for modeling such compositional vectors is the \emph{Dirichlet} distribution. In its basic form, a Dirichlet random vector \(\mathbf{x} = (x_1,\dots,x_K)\) is parameterized by a concentration vector \(\boldsymbol{\alpha} = (\alpha_1,\dots,\alpha_K)\) with \(\alpha_k>0\). The probability density function is
\[
p(\mathbf{x}\mid\boldsymbol{\alpha}) 
= \frac{1}{B(\boldsymbol{\alpha})} \prod_{k=1}^K x_k^{\alpha_k - 1}, 
\]
where 
\[
B(\boldsymbol{\alpha})
= \frac{\prod_{k=1}^K \Gamma(\alpha_k)}
       {\Gamma\Bigl(\sum_{k=1}^K \alpha_k\Bigr)},
\]
and \(\Gamma(\cdot)\) is the Gamma function. This parameterization captures both the support of compositional data (the simplex) and the potential correlation structure among components.

\subsection{B-DARMA Model}
To capture temporal dependence in compositional data, the Bayesian Dirichlet Auto-Regressive Moving Average (B-DARMA) model \citep{KATZ2024} augments the Dirichlet framework with VAR and VMA dynamics. Specifically, for each time \(t=1,\dots,T\), let 
\(\mathbf{y}_t\) be the observed composition. We assume
\begin{align}
    \mathbf{y}_t \;\bigm|\; \boldsymbol{\mu}_t,\;\phi_t &\sim \Dirichlet\bigl(\phi_t\,\boldsymbol{\mu}_t\bigr),
    \label{ytmodel}
\end{align}
with density $f(\bfy_t|\bfmu_t, \phi_t) \propto \prod_{j=1}^J y_{tj}^{\phi_t\mu_{tj}-1}$, where \(\boldsymbol{\mu}_t=(\mu_{t1},\dots,\mu_{tJ})\)' is the mean composition, and \(\phi_t>0\) is a precision parameter. Both \(\boldsymbol{\mu}_t\) and \(\phi_t\) may vary with time.

\paragraph{ALR link.} 
Because each \(\boldsymbol{\mu}_t\) lies in the simplex, we map it to an unconstrained \((J-1)\)-dimensional vector via the \emph{additive log-ratio} transform
\[
\alr\!\bigl(\boldsymbol{\mu}_t\bigr) \;=\; 
\Bigl(\ln\tfrac{\mu_{t1}}{\mu_{tJ}},\,\dots,\,\ln\tfrac{\mu_{t,J-1}}{\mu_{tJ}}\Bigr).
\]
We denote
\[
\bfeta_t \;=\; \alr\!\bigl(\boldsymbol{\mu}_t\bigr)\;\in\;\mathbb{R}^{J-1}.
\]

\paragraph{VARMA structure.}
To incorporate serial dependence, we assume \(\bfeta_t\) follows a vector VARMA process in the transformed space
\begin{align}
    \bfeta_t 
    \;=\; \sum_{p=1}^P \mathbf{A}_p \Bigl[\alr(\mathbf{y}_{t-p}) - \mathbf{X}_{t-p}\,\bfbeta\Bigr]
    \;+\; \sum_{q=1}^Q \mathbf{B}_q\!\Bigl[\alr(\mathbf{y}_{t-q}) - \bfeta_{t-q}\Bigr]
    \;+\; \mathbf{X}_t\,\bfbeta,
    \label{eq:eta_ARMA}
\end{align}
for \(t=m+1,\dots,T\), where \(m=\max(P,Q)\). In this notation
\begin{itemize}
    \item \(\mathbf{A}_p\) and \(\mathbf{B}_q\) are each \((J-1)\times(J-1)\) coefficient matrices.
    \item \(\mathbf{X}_t\) is a known \((J-1)\times r_\beta\) covariate matrix including any intercepts, trends, or seasonality.
    \item \(\bfbeta\in\mathbb{R}^{r_\beta}\) is a vector of regression coefficients shared across the \((J-1)\) components.
\end{itemize}

\paragraph{Precision parameter.} 
The Dirichlet precision \(\phi_t\) can also evolve over time. For an \(r_\gamma\)-vector of covariates \(\mathbf{z}_t\), we set
\begin{align}
    \phi_t \;=\; \exp\!\Bigl(\mathbf{z}_t\,\bfgamma\Bigr),
    \label{eq:phi_ARMA}
\end{align}
with \(\bfgamma\in\mathbb{R}^{r_\gamma}\). In the absence of covariates, we simply have \(\log \phi_t = \gamma\) for all \(t\), so \(\phi_t\) becomes a constant.

\paragraph{Parameter vector.}
We gather all unknown parameters into a vector of length $C$,
\[
\boldsymbol{\theta}
\;=\; \bigl(\,\mathbf{A}_{prs},\;\mathbf{B}_{qrs},\;\bfbeta',\;\bfgamma'\bigr)^\prime,
\]
where \(p=1,\dots,P\), \(q=1,\dots,Q\), and \(r,s=1,\dots,J-1\), and $\theta_j$ is the j-th element of $\bftheta$. The total number of free parameters is thus 
\[
C \;=\;(P+Q)\,(J-1)^2\;+\;r_\beta \;+\;r_\gamma.
\]
Bayesian inference begins by positing a prior distribution, $p(\bftheta)$, over the model parameters $\bftheta$. Bayes' theorem updates this prior to form the posterior
\begin{align*}
  p(\bftheta \mid \bfy_{1:T})
  \;=\;
  \frac{p(\bftheta)\, p(\bfy_{(m+1):T} \mid \bftheta, \bfy_{1:m})}
       {p(\bfy_{(m+1):T} \mid \bfy_{1:m})},
\end{align*}
where
\[
  p(\bfy_{(m+1):T} \mid \bftheta, \bfy_{1:m})
  \;=\;
  \prod_{t=m+1}^{T}
    p\bigl(\bfy_t \,\bigm|\,
        \bftheta,\,
        \bfy_{(t-m)!:,(t-1)}\bigr),
\]
and $p(\bfy_{(m+1):T}\mid \bfy_{1:m})$ is the normalizing constant obtained by integrating over $\bftheta$. Each $p(\bfy_t \mid \bftheta, \bfy_{(t-m):(t-1)})$ follows the Dirichlet likelihood \eqref{ytmodel}.

Next, to generate predictions for the subsequent $S$ time points, $\bfy_{(T+1):(T+S)}$, we construct the joint predictive distribution
\begin{align*}
  p\bigl(\bfy_{(T+1):(T+S)} \mid \bfy_{1:T}\bigr)
  \;=\;
  \int
    p\bigl(\bfy_{(T+1):(T+S)} \mid \bftheta\bigr)\,
    p\bigl(\bftheta \mid \bfy_{1:T}\bigr)
  \, d\bftheta.
\end{align*}
In practice, analysts often summarize this distribution at future time points $t \in (T+1):(T+S)$ by reporting measures such as the posterior mean or median.

\subsection{Bayesian Shrinkage Priors for B-DARMA Coefficients}
\label{sec:priors}

In a fully Bayesian approach, all unknown parameters in the B-DARMA model --- including the VARMA coefficients in \(\mathbf{A}_p\) and \(\mathbf{B}_q\), the regression vector \(\boldsymbol{\beta}\), and the precision-related parameters \(\boldsymbol{\gamma}\) --- require prior distributions. Different shrinkage priors can produce significantly different outcomes, particularly in high-dimensional or sparse settings where many coefficients may be small. This section discusses five popular priors (normal, horseshoe, Laplace, spike-and-slab, and hierarchical), highlighting how each encodes shrinkage or sparsity assumptions.

An \emph{informative normal prior} serves as a straightforward baseline. We model each coefficient \(\theta_j\) by \(\theta_j \sim \mathcal{N}(a,b^2)\). The mean \(a\) often defaults to zero unless prior knowledge indicates a different center. The variance \(b^2\) determines shrinkage strength; smaller \(b^2\) yields tighter concentration around \(a\). In B-DARMA applications, it may be sensible to set \(b=1\) for the VARMA parameters if they are believed to be small on average, whereas for covariates \(\boldsymbol{\beta}\), a smaller prior variance such as \(b^2=0.01\) can reflect stronger beliefs that regression effects are modest.

The \emph{horseshoe prior} \citep{Carvalho2010} is well-suited for sparse problems. We model each coefficient \(\nu\) as \(\nu \mid \tau,\lambda_\nu \sim \mathcal{N}(0,\tau^2 \lambda_\nu^2)\) with a global scale \(\tau\sim \mathrm{Cauchy}_+(0,1)\) and local scales \(\lambda_\nu\sim \mathrm{Cauchy}_+(0,1)\). Large local scales allow some coefficients to remain sizable, whereas most are heavily shrunk. This can be beneficial if the B-DARMA model includes many possible lags or covariates, only a small subset of which are expected to matter for compositional forecasting.

A \emph{Laplace (double-exponential) prior} \citep{Park2008} employs the density \(p(\nu\mid b)=\frac{1}{2b}\exp(-|\nu|/b)\). This enforces an \(\ell_1\)-type penalty that can drive many coefficients close to zero while still allowing moderate signals to persist. The scale \(b\) can be chosen a priori or assigned its own hyperprior, such as a half-Cauchy, so that the data adaptively determine the shrinkage level. In a B-DARMA context, choosing a smaller \(b\) for high-dimensional VARMA terms can prevent spurious estimates from inflating the parameter space.

A \emph{spike-and-slab prior} \citep{Mitchell1988} can introduce explicit sparsity by placing a point mass at zero. 
Let $\tau_j \sim \mathrm{Beta}(1,1)$ be the mixing parameter for the $j$th coefficient. 
Then each coefficient $\theta_j$ follows the mixture
\[
\theta_j \;\sim\; \tau_j\,\mathcal{N}(0,1)\;+\;(1-\tau_j)\,\delta_0,
\]
where $\delta_0$ is the Dirac measure at zero. 
Coefficients drawn from the spike component remain exactly zero, effectively excluding them from the model, 
while coefficients from the slab remain freely estimated. 
This setup allows the B-DARMA specification to adapt by discarding irrelevant lags or covariates.

A \emph{hierarchical shrinkage prior} \citep{Polson2012} can encourage partial pooling across groups of coefficients. We model each coefficient \(\nu\) by \(\nu\mid\sigma\sim \mathcal{N}(0,\sigma^2)\) and then place a half-Cauchy prior on \(\sigma\). In B-DARMA, one could assign separate group scales to the AR, MA, and regression blocks, thereby allowing the model to learn an appropriate overall variability for each group of parameters. 

These priors differ in how strongly they push coefficients toward zero and whether they favor a few large coefficients or moderate shrinkage for all. The normal prior provides a baseline continuous shrinkage, the horseshoe excels when only a minority of parameters are truly large, the Laplace induces an \(\ell_1\)-type penalty that can zero out many coefficients, spike-and-slab explicitly discards some parameters, and the hierarchical alternative enables group-level learning of shrinkage scales. The next sections illustrate how these choices affect both parameter recovery and compositional forecasting in various simulation and real-data scenarios.

\subsection{Posterior computation}\label{sec:postcomp}

All B-DARMA models are fit with STAN \citep{Rstan} in R using Hamiltonian Monte Carlo. We run \(4\) chains, each with \(500\) warm-up and \(750\) sampling iterations, yielding \(3{,}000\) posterior draws. The sampler uses \texttt{adapt\_delta}\,=\,0.85, \texttt{max\_treedepth}\,=\,11, and random initial values drawn uniformly from \([-.25,.25]\).

\subsection{Stationarity and structural considerations}
\label{sec:stationarity}
\paragraph{Open theoretical gap.}
B-DARMA inherits the difficulty noted by \citet{zheng2017dirichlet}: 
after the additive-log-ratio (ALR) link the innovation is \emph{not} a
martingale-difference sequence (MDS).  Consequently, classical
stationarity results for VARMA\,(p,q) processes do not transfer
directly, and a full set of strict- or weak-stationarity conditions
remains an open problem \citep[IJF, §2.2]{KATZ2024}.

\paragraph{Why Bayesian inference remains valid.}
Bayesian estimation proceeds via the joint posterior
\(p(\bftheta\mid\bfy_{1:T})\propto p(\bfy_{1:T}\mid\bftheta)\,p(\bftheta)\),
which does not require the likelihood-error sequence to be an MDS
\citep{gelman2020bayesian}.  We therefore regard
B--DARMA as a flexible likelihood-based filter for
compositional time series rather than assume the fitted parameters
represent a strictly stationary data-generating process.

\section{Simulation Studies}
\label{sec:simulation}

We conduct three simulation studies to investigate how different priors affect parameter inference and forecasting in a B-DARMA model. All studies use the same sparse DARMA(2,1) data-generating process (DGP) but vary the fitted model to be correctly specified, deliberately overfitted, or underfitted. We first describe the DGP and the priors considered, then outline the study designs and performance metrics.

\subsection{Data-Generating Process}
\label{subsec:dgp}

We simulate a six-dimensional compositional time series \(\mathbf{y}_t\), each component constrained to sum to one. The true process is DARMA(2,1) with fixed precision \(\phi=500\), \(\boldsymbol{\beta} = (0.1,\,-0.05,\,0.03,\,-0.02,\,0.04)^\prime\), and $\bfX_t=I_5$, a $5 \times 5$ identity matrix. We set our VAR and VMA matrices to, 
\[
\mathbf{A}_1 = 
\begin{bmatrix}
 0.80 &  0.05 & -0.04 & -0.05 & -0.05 \\
-0.01 &  0.70 & -0.03 &  0.02 & -0.01 \\
 0.02 &  0.00 &  0.90 &  0.02 &  0.04 \\
-0.03 & -0.07 & -0.02 &  0.85 & -0.01 \\
 0.04 & -0.02 &  0.01 & -0.01 &  0.75 
\end{bmatrix}
\quad
\mathbf{A}_2 = 
\begin{bmatrix}
-0.30 &  0.03 &  0.02 &  0.05 & -0.04 \\
 0.02 & -0.20 & -0.01 & -0.02 &  0.01 \\
-0.01 &  0.05 & -0.25 & -0.01 &  0.01 \\
-0.01 &  0.04 &  0.01 & -0.15 &  0.00 \\
 0.06 &  0.00 & -0.11 & -0.02 & -0.20
\end{bmatrix},
\]
\[
\mathbf{B}_1 = 
\begin{bmatrix}
 0.50 & -0.02 &  0.03 &  0.00 &  0.03 \\
 0.05 &  0.40 &  0.03 & -0.01 &  0.02 \\
 0.02 &  0.01 &  0.45 & -0.02 &  0.13 \\
-0.01 &  0.10 &  0.05 &  0.35 &  0.01 \\
-0.01 &  0.04 & -0.11 &  0.10 &  0.40
\end{bmatrix}.
\]
 We initialize \(\mathbf{y}_1\) and \(\mathbf{y}_2\) from \(\mathrm{Dirichlet}(\mathbf{1})\), simulate \(T=100\) observations, and replicate this procedure 50 times. We focus on posterior inference for \(\mathbf{A}_p,\;\mathbf{B}_q,\) and \(\boldsymbol{\beta}\), as well as forecasting accuracy.

\subsection{Prior Distributions and Hyperparameters}
\label{subsec:sim_priors}

We specify five candidate priors for the coefficient vectors: Normal (mean 0, variance 1), Horseshoe (global and local $\mathrm{Cauchy}_+(0,1)$ scales), Laplace (scale $b=1$), spike-and-slab (with local mixing parameters $\tau_j \sim \mathrm{Beta}(1,1)$), and hierarchical Normal (with a half-Cauchy prior on the group-level scale). Although these priors remain the same across simulations, they are applied to different sets of coefficients: $\mathbf{A}_1,\mathbf{A}_2,\mathbf{B}_1,$ and $\boldsymbol{\beta}$ in Simulation~1; $\mathbf{A}_1$--$\mathbf{A}_4,\mathbf{B}_1$--$\mathbf{B}_2,$ and $\boldsymbol{\beta}$ in Simulation~2; and $\mathbf{A}_1$ and $\boldsymbol{\beta}$ in Simulation~3. For the Dirichlet precision parameter, we use $\gamma_\phi \sim \mathcal{N}(7,1.5)$. The Normal prior on $\boldsymbol{\beta}$ is $\mathcal{N}(0,0.1)$, and the Laplace and hierarchical Normal versions adopt smaller scales to impose stronger shrinkage on intercepts.

\subsection{Study Designs}
\label{sec:study_designs}

We fit three B-DARMA specifications to the same DARMA(2,1) data
\begin{itemize}
    \item \textbf{Study 1 (Correct Specification).} B-DARMA(2,1) matches the true DGP.
    \item \textbf{Study 2 (Overfitting).} B-DARMA(4,2) includes extraneous higher-order VAR and VMA terms.
    \item \textbf{Study 3 (Underfitting).} B-DARMA(1,0) omits the second VAR lag and the MA(1) term.
\end{itemize}
Each configuration is paired with each of the five priors, yielding 15 total fitted models. We repeat the simulation for 50 synthetic datasets of length \(T\).

\subsection{Evaluation Metrics}
\label{subsec:eval_metrics}

We assess both parameter recovery and forecasting performance. Let \(\theta_{\text{true}}\) be a true parameter and \(\hat{\theta}^{(s)}\), the posterior mean from simulation \(s\). We compute
\[
\mathrm{Bias}_j 
= \frac{1}{50}\sum_{s=1}^{50} \bigl(\hat{\theta}_j^{(s)} - \theta_{j,\text{true}}\bigr),
\quad
\mathrm{RMSE}_j
=\sqrt{ \frac{1}{50}\sum_{s=1}^{50}\bigl(\hat{\theta}_j^{(s)} - \theta_{j,\text{true}}\bigr)^2},
\]
along with 95\% credible-interval coverage and interval length. If a parameter is omitted (as in underfitting), we exclude it from these summaries.

For forecasting, we use the first \(80\) points for training and the remaining \(20\) points for testing. Let \(\hat{\mathbf{y}}_{t}^{(s)}\) be the posterior mean forecast at time \(t\) in simulation \(s\). We define
\[
\mathrm{RMSE}_{\text{forecast}}
= \sqrt{
  \frac{1}{50 \times 20 \times 6}
  \sum_{s=1}^{S} 
    \sum_{t=1}^{20}
      \sum_{k=1}^{6}
        \bigl(y_{t,k}^{(s)} \;-\; \hat{y}_{t,k}^{(s)}\bigr)^2
},
\]
where \(\mathbf{y}_{t}^{(s)}\) denotes the true composition in the test set for the $s$-th simulation. Each of the three study designs is then evaluated under each of the five priors, illuminating how prior choice interacts with model misspecification.

\section{Results}

The summarized parameter estimation results of the three simulation studies are shown in Tables~\ref{tab:parameter_summary}-\ref{tab:parameter_summary_3} and the summarized forecast results are shown in Tables \ref{tab:forecast_summary_combined}-\ref{tab:forecast_summary_within_ratios}. 

\subsection{Study 1: Correct Model Specification (DARMA(2,1))}
\label{sec:study1}

\subsubsection{Parameter Estimation}

Table~\ref{tab:parameter_summary} includes results for the correctly specified DARMA(2,1). All priors yield estimates that align with the true parameter values, and both bias and RMSE remain modest. Certain shrinkage priors, including the Horseshoe and Laplace, produce slightly lower RMSE and narrower intervals than the normal prior. Hierarchical priors also reduce parameter uncertainty to some extent, although the differences relative to other shrinkage methods are not pronounced. Under spike-and-slab, small but non-zero signals can be set to zero, occasionally lowering coverage (for example, a coverage rate of 0.8240 for the intercept \(\beta\)).

\subsubsection{Forecasting Performance}

In Table~\ref{tab:forecast_summary_combined} (Sim~1 columns), the average forecast RMSE spans 0.031--0.032, with minimal variation among priors. The Horseshoe, Laplace, and Hierarchical prior provide slight gains in predictive accuracy compared to the Informative Normal or Spike-and-Slab approaches.

\subsection{Study 2: Overfitting Scenario (B-DARMA(4,2))}
\label{sec:study2}

\subsubsection{Parameter Estimation}

When B-DARMA(4,2) is applied to data generated by a DARMA(2,1) process, additional VAR and VMA terms that should be zero are introduced. In Table \ref{tab:parameter_summary_2}, relatively large biases and inflated RMSE values appear under the informative normal prior for these extraneous coefficients. Horseshoe priors shrink many of those coefficients near zero, reflected in low RMSE (for instance, 0.0176 or 0.0161 for \(A_3\) and \(A_4\)) and coverage rates close to the nominal level. Spike-and-slab also suppresses unwanted terms, although borderline signals may be excessively reduced.

\subsubsection{Forecasting Performance}

Forecast RMSE under overfitting conditions is listed in Table~\ref{tab:forecast_summary_combined} (Sim~2 columns). Higher errors (0.0341) are observed for the Informative Normal prior, while the Horseshoe produces the lowest RMSE (0.0315). Laplace and Hierarchical approaches also outperform the weakly regularized alternative. Ratios in Table~\ref{tab:forecast_summary_ratios_expanded} indicate that Horseshoe’s forecast RMSE in the overfitted model differs little from the correctly specified case.

\subsection{Study 3: Underfitting Scenario (B-DARMA(1,0))}
\label{sec:study3}

\subsubsection{Parameter Estimation}

In Table~\ref{tab:parameter_summary_3}, broad increases in RMSE and reduced coverage are noted when the second VAR lag and the VMA(1) term are omitted, regardless of the prior used. Horseshoe, Laplace, and Spike-and-Slab cannot compensate for missing structural components, which leads to biased AR(1) estimates and coverage gaps. The hierarchical prior exhibits comparable issues.

\subsubsection{Forecasting Performance}

Forecast RMSE remains near 0.032--0.033, indicating a 3--4\% rise over the correct specification. This shortfall is consistent across priors (Table~\ref{tab:forecast_summary_ratios_expanded}) confirming that underspecification cannot be resolved by shrinkage.

\subsection{Overview of Results}

These simulations illustrate how model specification and prior choice jointly affect B-DARMA outcomes. Under a correctly specified DARMA(2,1), all priors capture the process well, although Horseshoe and Hierarchical priors yield slightly lower RMSE and narrower intervals. In overfitted scenarios, shrinkage priors—especially Horseshoe—readily suppress spurious parameters and maintain robust forecasts. Underfitting cannot be mitigated by any prior, as exclusion of critical VAR or VMA terms inflates bias and diminishes coverage. Further exploration of alternative coefficient matrices, described in the Supplementary Material, reinforces these findings: shrinkage priors protect against overfitting but cannot repair underspecified models. Horseshoe, Laplace, and Hierarchical priors yield stable forecasts across various parameter settings, whereas the Informative Normal approach is more susceptible to inflated estimates in large or sparse models.

\section{Application to S\&P 500 Sector Trading Values}
\label{application}

\subsection{Motivation and Data Description}

A fixed daily trading value in the S\&P~500 is allocated across different sectors, giving rise to compositional data that captures how investors distribute their capital each day. Tracking these proportions over time can reveal macroeconomic trends, sector rotation, and shifts in investor sentiment. In our analysis, we examined the daily proportions of eleven S\&P~500 sectors from January~2021 through December~2023. These sectors include:

\begin{itemize}
  \item \textbf{Technology}: software, hardware, and related services
  \item \textbf{Healthcare}: pharmaceuticals, biotechnology, and healthcare services
  \item \textbf{Financials}: banking, insurance, and investment services
  \item \textbf{Consumer Discretionary}: non-essential goods and services
  \item \textbf{Industrials}: manufacturing, aerospace, defense, and machinery
  \item \textbf{Consumer Staples}: essential goods, such as food and household items
  \item \textbf{Energy}: oil, gas, and renewable resources
  \item \textbf{Utilities}: public services such as electricity and water
  \item \textbf{Real Estate}: REITs and property management
  \item \textbf{Materials}: chemicals, metals, and construction materials
  \item \textbf{Communication Services}: media, telecommunication, and internet services
\end{itemize}

Let \(V_{kt}\) denote the dollar trading value executed in sector \(k\)
on trading day \(t\) for \(k=1,\dots,K\) (\(K=11\) sectors).  
The market‐wide total is  
\[
  g_t \;=\; \sum_{k=1}^{K} V_{kt},
\]
and the composition we analyse is the vector of sector shares  
\[
  \mathbf{y}_t \;=\; (y_{1t},\dots,y_{Kt})^{\mathsf T}, 
  \qquad y_{kt}=V_{kt}/g_t,
  \qquad \sum_{k=1}^{K} y_{kt}=1.
\]
The stationarity (or non-stationarity) of the compositional time series
\(\mathbf{y}_t\) is independent of that of the gross total \(g_t\):
\(g_t\) may drift without affecting the stationarity of the shares,
and, conversely, the shares could be non-stationary even if \(g_t\) is itself stationary.

\begin{figure}[ht]
    \centering
    \includegraphics[width=1\textwidth]{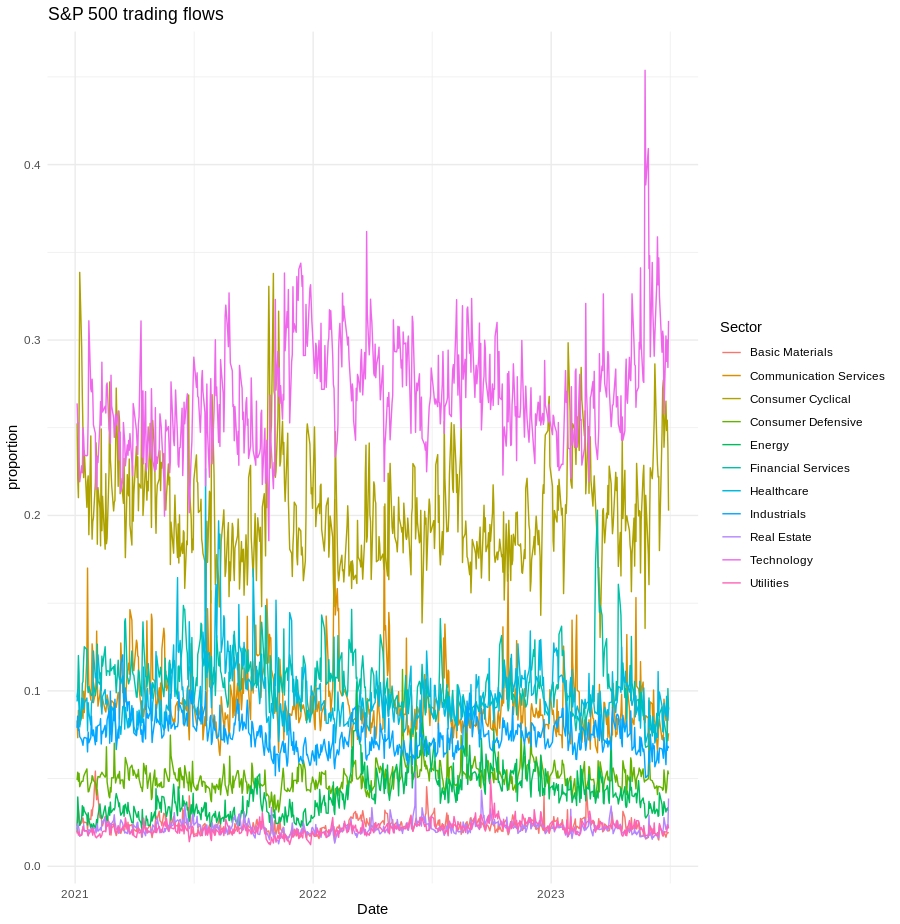} 
    \caption{Time series of S\&P 500 trading flows by sector from January~2021 to June~2023.  Technology (magenta) and Financial Services (goldenrod) are consistently among the largest proportions.}
    \label{fig:sp500_2021}
\end{figure}

We used January~1, 2021, to June~30, 2023, as our training data as shown in figure \ref{fig:sp500_2021}. Forecast evaluation was conducted on the following 126 trading days, spanning July~1 through December~31, 2023.

\begin{figure}[ht]
    \centering
    \includegraphics[width=0.7\textwidth]{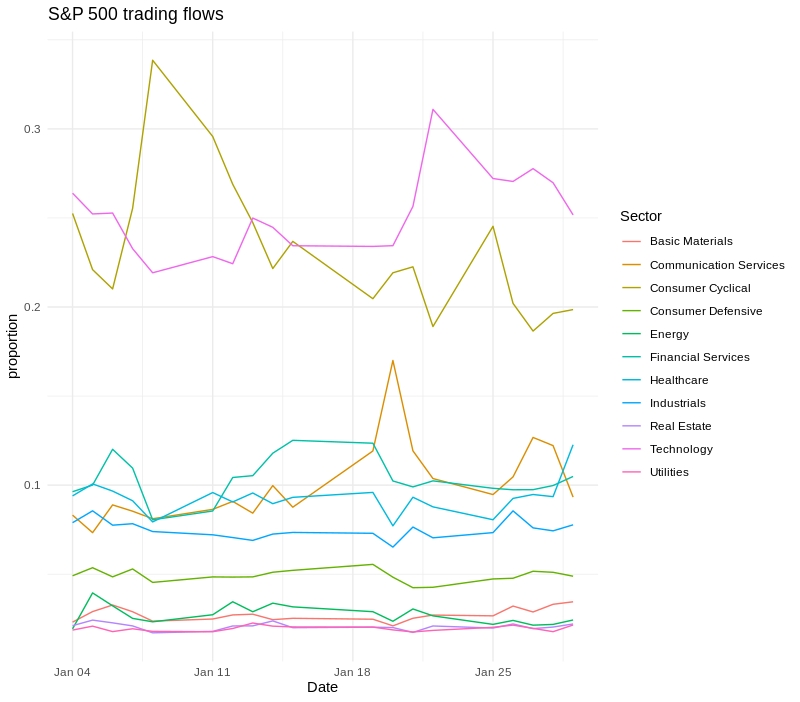} 
    \caption{Daily S\&P 500 trading flows by sector for January~2021.  The shortened time window highlights intramonth volatility, especially for the Consumer Cyclical (orange) and Communication Services (yellow) sectors.}
    \label{fig:sp500_jan}
\end{figure}

\begin{figure}[ht]
    \centering
    \includegraphics[width=1\textwidth]{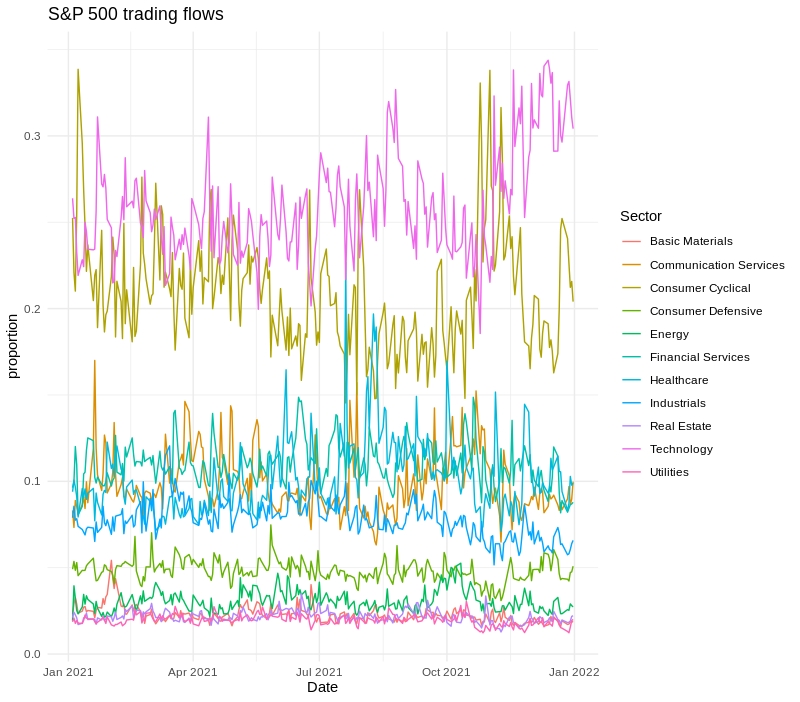} 
    \caption{Time series of S\&P 500 trading flows from january 2021 to Jan 2022.  Despite daily and yearly fluctuations, Technology remains among the top in proportion of total trading value.}
    \label{fig:sp500_long}
\end{figure}

\begin{figure}[ht]
    \centering
    \includegraphics[width=01\textwidth]{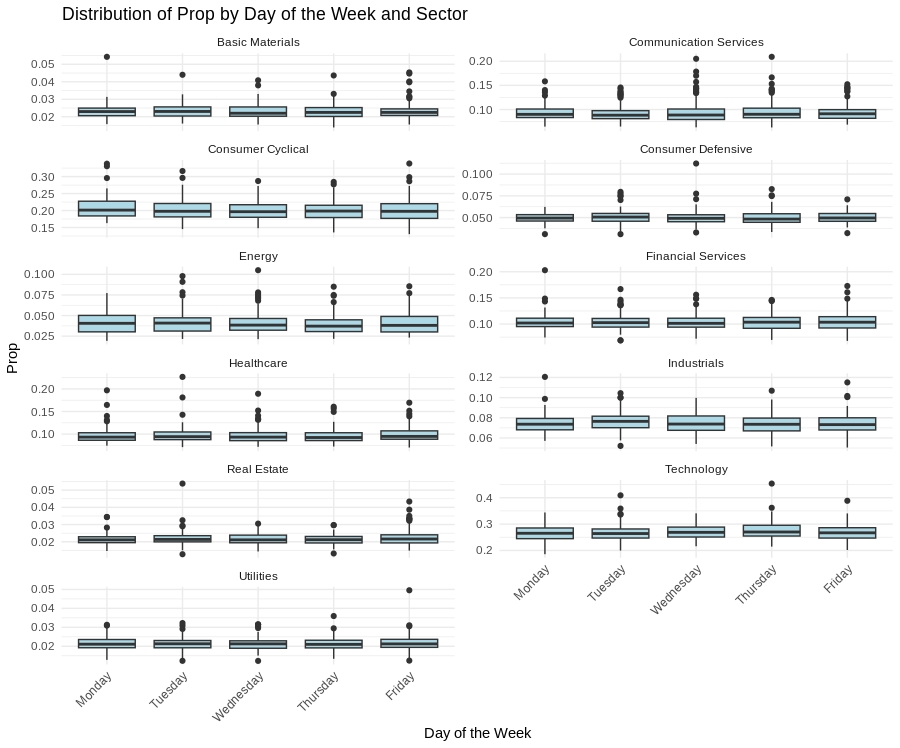} 
    \caption{Boxplots of S\&P 500 sector trading proportions by day of the week.  Outliers primarily occur for Consumer Cyclical and Technology, though smaller sectors also display occasional spikes.}
    \label{fig:sp500_boxplot}
\end{figure}

Short-term variability, cyclical tendencies, and gradual long-term shifts emerge across the series. Technology and Financials consistently hold the largest fractions of the fixed daily trading value, whereas Utilities and Real Estate occupy much smaller shares. Day-to-day fluctuations tend to be more pronounced in Consumer Discretionary and Communication Services, indicating greater sensitivity to rapidly changing market conditions. Seasonal patterns and moderate differences across weekdays are also evident, as illustrated by Figures~\ref{fig:sp500_jan}--\ref{fig:sp500_boxplot}, further underscoring the multifaceted nature of sector-level trading dynamics.

\paragraph{Model specification (purposeful over-parameterisation).}
A \(\text{B--DARMA}(P{=}10,Q{=}0)\) model is fitted \emph{by design} with a lag order that exceeds the horizon over which sector reallocations are generally thought to propagate.  Ten trading days (\(\approx\) two calendar weeks) therefore represent a deliberate over-fit, allowing us to evaluate how strongly the global–local shrinkage priors suppress redundant dynamics.  Each sector’s additive-log-ratio is regressed on its own composition at lags \(1,\dots,10\), yielding \(1{,}000\) VAR coefficients in the matrices \(A_{1},\dots,A_{10}\).

Seasonality is modelled with Fourier bases that are \emph{fixed} regardless of the chosen lag: two sine–cosine pairs capture the 5-day trading cycle (Monday–Friday) and five pairs capture the annual cycle of roughly 252 trading days.  These 14 terms, together with a sector-specific intercept, form the design matrix for the linear predictor; the same seasonal structure enters the model for the Dirichlet precision \(\phi_t\).

In total, the specification estimates \(1{,}165\) parameters: \(1{,}000\) VAR coefficients, \(140\) seasonal coefficients, \(10\) intercepts, and \(15\) precision-related terms.  The intentionally generous lag order thus functions as a stress test.

\subsection{Priors and Hyperparameters}

For all priors, the intercept in $\gamma_\phi$ is given a $\mathcal{N}(7,1.5)$ prior, and all seasonal Fourier terms in $\gamma_\phi$ receive $\mathcal{N}(0,0.1)$. 

Under the \emph{informative normal} prior, every element of $\boldsymbol{\beta}$ is modeled with a $\mathcal{N}(0,0.1)$ prior, and each element of $\mathbf{A}_p$ has a $\mathcal{N}(0,1)$ prior. 

The \emph{Laplace} prior \citep{Park2008} is implemented via an exponential mixture: each coefficient $\theta_i$ satisfies $\theta_i \sim \mathcal{N}(0,\nu)$ with $\nu \sim \mathrm{Exponential}(1/b)$, leading to an overall Laplace$(0,b)$ prior. We set $b_\beta=1.0$ and $b_A=1.0$ for $\boldsymbol{\beta}$ and $\mathbf{A}_p$ respectively.

For the \emph{spike-and-slab} prior \citep{Mitchell1988}, each coefficient $\theta_i$ is drawn from a mixture $\theta_i = 0$ with probability $(1-\tau_i)$ or $\theta_i \sim \mathcal{N}(0,1)$ with probability $\tau_i$, where $\tau_i \sim \mathrm{Beta}(1,1)$ is the mixing weight.

The \emph{horseshoe} prior \citep{Carvalho2010} introduces one global scale parameter $\tau$ and local scales $\lambda_i$, all drawn from $\mathrm{Cauchy}^+(0,1)$. Each coefficient $\theta_i$ then follows $\mathcal{N}(0,\,\tau\,\lambda_i)$.

Finally, under the \emph{hierarchical} prior  \citep{Polson2012}, the coefficients are partitioned into three groups: 
(i) the elements of $\boldsymbol{\beta}$, 
(ii) the diagonal entries of each lag matrix $\mathbf{A}_p$, 
and (iii) the off-diagonal entries of each $\mathbf{A}_p$. 
Then the coefficients in $\boldsymbol{\beta}$ are modeled as 
$\beta_i \,\mid\, \sigma_{\beta} \sim \mathcal{N}(0,\sigma_\beta)$, 
while for each lag $p$ in $\mathbf{A}_p$, the diagonal entries follow 
$\mathcal{N}(0.5,\sigma_A)$ and the off-diagonal entries follow 
$\mathcal{N}(0,\sigma_{A,\text{off}})$.

\subsection{Evaluation Metrics and Forecasting}

A B-DARMA(10,0) model was fit with each of the five priors using the training data. A 126 day forecast was then generated for the test period using the mean of the joint predictive distribution. Mean absolute error (MAE) was used as a measure of average discrepancy between predicted and observed shares, while root mean squared error (RMSE) weighs larger deviations more explicitly.

\subsection{Results}

\begin{figure}[!ht]
\centering
\includegraphics[width=1.15\textwidth]{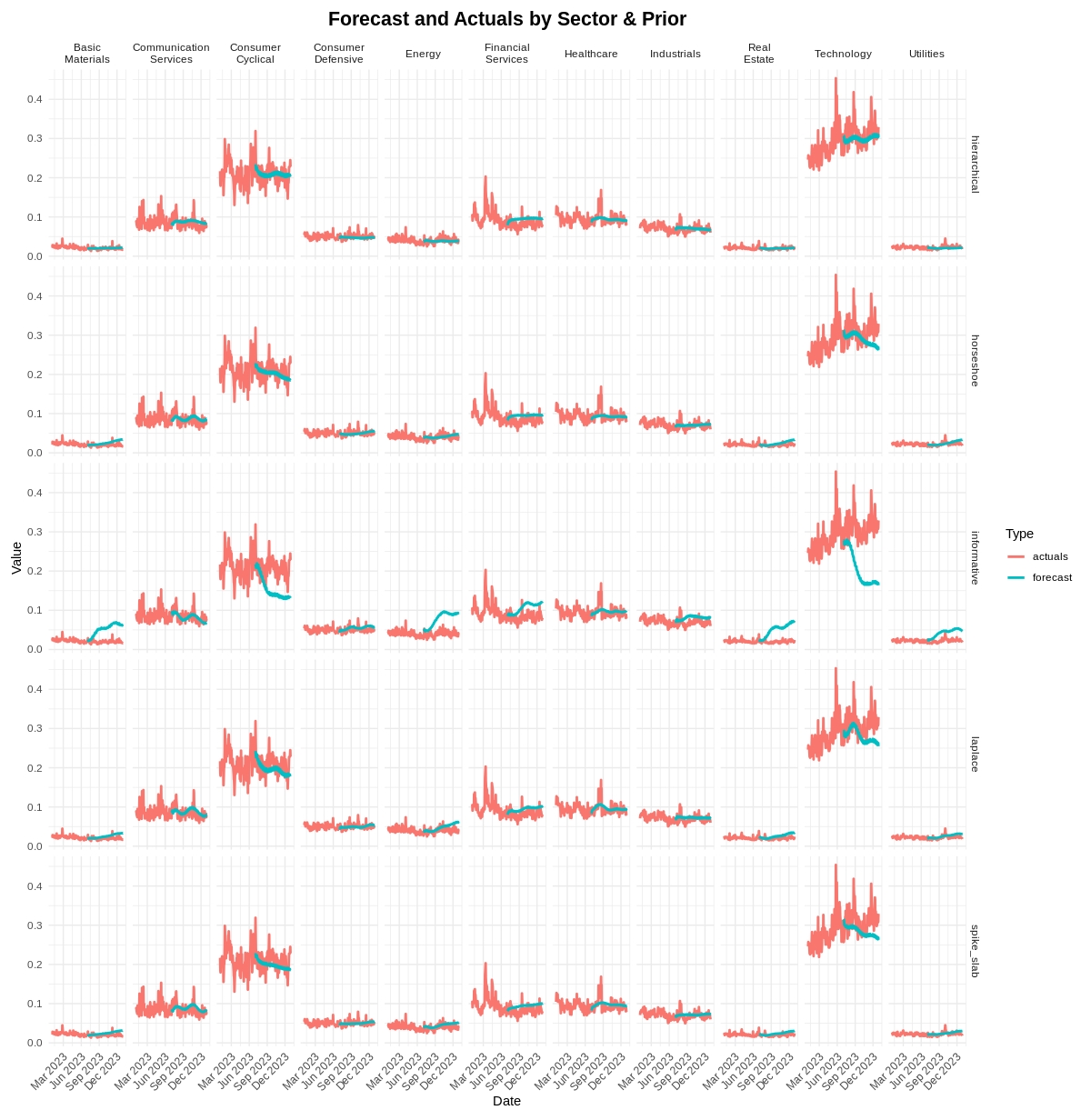}
\caption{Forecast and Actuals by Sector and Prior for S\&P 500 Data. Red lines indicate actual daily sector proportions; turquoise lines are the forecasts from the fitted B-DARMA model. Horseshoe and hierarchical priors tend to track actual series more tightly than the informative prior in many sectors.}
\label{fig:forecast_facet}
\end{figure}

In Figure~\ref{fig:forecast_facet}, forecasts (in turquoise) are shown together with actual daily proportions (in red) for each sector over the 126-day test interval. Hierarchical and Horseshoe priors led to predictions that aligned more closely with the observed values compared to the informative prior, particularly in volatile sectors such as Energy and Technology. Under the Spike-and-Slab prior, smaller signals were at times set to zero, which occasionally delayed responses to abrupt shifts in sector composition.

\begin{figure}[!ht]
\centering
\includegraphics[width=01\textwidth]{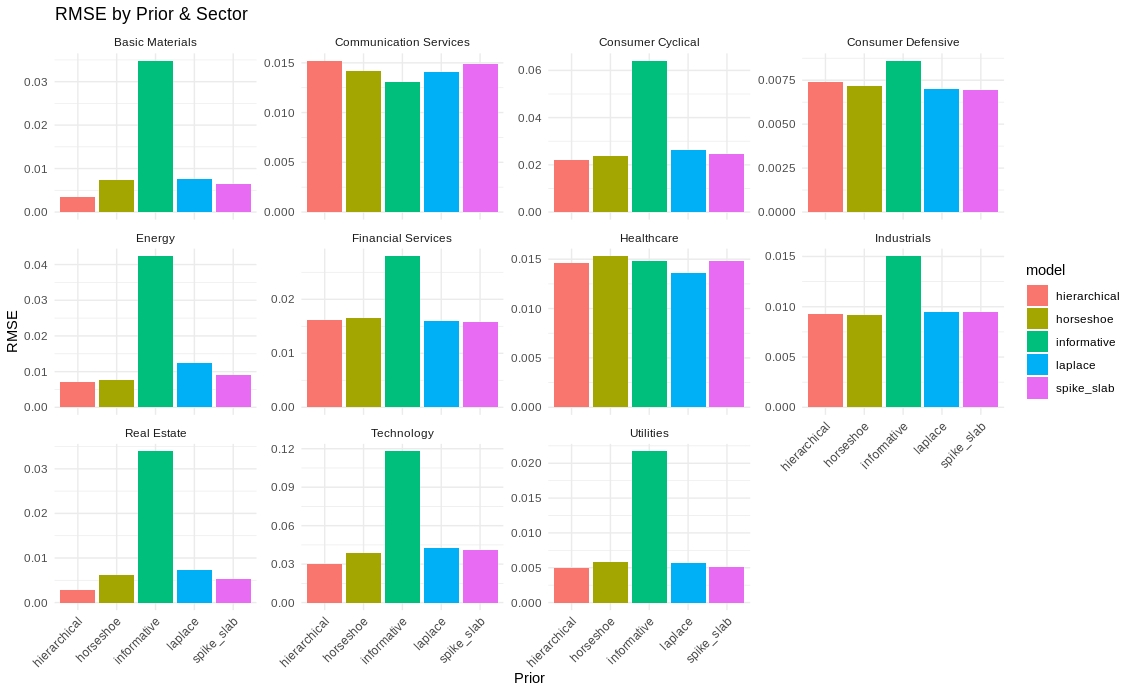}
\caption{RMSE by Prior and Sector for S\&P 500 forecasting. Each facet shows one sector’s RMSE across the five priors. Shrinkage priors (horseshoe, laplace, spike\_slab) and hierarchical strategies generally outperform the more loosely regularized informative prior.}
\label{fig:rmse_barplots}
\end{figure}

Sector-level RMSE values under each prior appear in Figure~\ref{fig:rmse_barplots}, and average RMSE and MAE are listed in Table~\ref{tab:error_table_summary}. Hierarchical and spike-and-slab priors yielded the smallest RMSE in multiple sectors, and Horseshoe priors were found to be effective for larger fluctuations in Consumer Cyclical and Financial Services. The informative prior generally had higher errors, reflecting the value of regularization in a high-dimensional parameter space.

\subsubsection{Summary of Findings}
The findings in this real-data application align with our earlier simulation results. In the large-scale B-DARMA model, shrinkage priors reduced extraneous complexity and lowered forecast errors. Horseshoe, Laplace, and spike-and-slab priors limited the number of active coefficients and delivered stable predictions. Hierarchical priors performed similarly, in part because they \emph{partially pool} information across related parameters—shrinking them toward a common distribution while still allowing differences where the data support them. Sectors with higher volatility, such as Technology, Communication Services, and Consumer Cyclical, appeared to benefit most from strong shrinkage, whereas more stable sectors like Utilities and Basic Materials performed comparably under all priors. Overall, these outcomes highlight the value of effective regularization in models with multiple lags and complex seasonality, especially in markets that experience rapid fluctuations.

\section{Discussion}

\subsection{Comparisons Across Simulations and Real Data}

Our three simulation studies demonstrate how prior selection interacts with model specification in B-DARMA. When the model order matches the true data-generating process (DARMA(2,1)), all priors yield acceptable results, though Horseshoe and Hierarchical priors produce slightly lower RMSE and better interval coverage. Overfitting by fitting B-DARMA(4,2) confirms that strong shrinkage—particularly Horseshoe—helps suppress spurious higher-order coefficients and avoids inflating forecast errors. Conversely, underfitting remains impervious to prior choice, as omitting critical AR(2) and MA(1) terms leads to uniformly higher biases and coverage shortfalls across all priors. All four shrinkage priors neutralise redundant lags, and any of them is preferable to an under-specified alternative.

These lessons translate directly to real data, where we intentionally specified a large-lag B-DARMA model for S\&P 500 sector trading. Just as in the overfitting simulation, Horseshoe, Laplace, and Spike-and-Slab priors effectively shrank extraneous parameters and preserved forecast accuracy. Hierarchical priors provided comparably strong performance through partial pooling, whereas the nformative prior yielded higher forecast errors in volatile sectors such as Energy and Technology. These empirical patterns mirror the simulation findings, reinforcing that the combination of a potentially over-parameterized model and minimal shrinkage can lead to unstable estimates and suboptimal forecasts.

\paragraph{Computation time.}
Using identical \texttt{Stan} settings (4 chains, 1 250 iterations, 500 warm-up), the informative prior completed fastest in about 10 min per chain and serves as our baseline.  Laplace added roughly 20 \% to wall-clock time (12 min), reflecting the heavier double-exponential tails that require more leapfrog steps.  Horseshoe and Hierarchical priors were slower by 40 \% (14–15 min) because their global–local scale hierarchies force smaller step sizes during HMC integration.  Spike-and-Slab was the clear outlier, nearly doubling run-time (20 min) owing to the latent inclusion indicators that create funnel-shaped geometry. Practitioners with tight compute budgets might prefer the Laplace prior, while more aggressive sparsity (Horseshoe, Hierarchical, Spike-and-Slab) simply warrants proportionally longer runs.

\subsection{Implications and Guidelines}

Overall, our results underscore three main points relevant to both simulated and real-world compositional time-series modeling:

\begin{itemize}
    \item \textbf{Shrinkage priors mitigate overfitting.} Horseshoe priors are especially adept at handling sparse dynamics and large parameter spaces, as shown by the minimal performance degradation in both simulated and real overfitting scenarios.
    \item \textbf{Hierarchical priors offer robust partial pooling.} They attain performance comparable to Horseshoe while retaining smooth shrinkage across correlated parameters, making them a flexible choice for multi-sector or multi-component series.
    \item \textbf{Model mis-specification overshadows prior advantages.} Underfitting in simulations, or failing to include essential lags in practice, leads to systematic bias and reduced coverage that shrinkage alone cannot remedy. Model identification and appropriate lag selection remain critical for accurate inference.
\end{itemize}

The S\&P 500 data analysis underscores these points: shrinkage priors consistently outperformed the informative prior in managing a high-dimensional model, yet sector-specific volatility still caused reduced accuracy. Sectors with greater day-to-day variability, such as Technology or Consumer Cyclical, benefited more from aggressive regularization than stable ones (e.g., Utilities, Basic Materials).

Incorporating dynamic selection of lag structures or exploring alternative priors may further improve compositional forecasting in settings with complex seasonalities or extremely large parameter spaces. Meanwhile, practitioners should pair robust prior modeling with careful model diagnostics (e.g., residual checks, information criteria) to ensure that underfitting and overfitting are detected early.

Overall, these findings reinforce the need for carefully chosen priors—especially in high-dimensional compositional time-series contexts. Horseshoe, Laplace, and Hierarchical priors successfully protect against overfitting while preserving essential signals, as illustrated by both simulation and real-data (S\&P 500) analyses. Nonetheless, fundamental model adequacy remains paramount, as even the best prior cannot rescue a structurally under-specified model. We recommend that analysts combine thorough sensitivity checks of priors with well-informed decisions about lag order, covariate inclusion, and potential seasonal effects in B-DARMA applications.

\subsection*{Code Availability}
All \texttt{R} scripts and \texttt{Stan} model files used in this study are publicly available at \\
\href{https://github.com/harrisonekatz/bdarma_sensitivity_analysis}{https://github.com/harrisonekatz/bdarma\_sensitivity\_analysis}. 
All results and figures in this manuscript can be reproduced by running the scripts found in that repository.

\section*{Conflicts of Interest}
The authors declare no conflicts of interest and that all work and opinions are their own and that the work is not sponsored or endorsed by Airbnb.

\section*{Tables \& Figures}

\begin{table}[ht]
\centering
\caption{\textbf{Parameter Estimation Summary for Simulation Study 1 (Correct Specification).}
We show mean bias, RMSE, average credible interval length, and coverage for each prior, summarizing over all parameters in each of $\beta, A_1, A_2,$ and $B_1$. Lower RMSE and shorter intervals typically indicate more effective shrinkage, while coverage near the nominal 0.95 is desirable.}
\begin{tabular}{lccccc}
\hline
\textbf{Coefficient} & \textbf{Prior}      & \textbf{Mean Bias} & \textbf{Mean RMSE} & \textbf{Mean CI Length} & \textbf{Coverage} \\
\hline
\multirow{5}{*}{$\beta$}
& Informative   & -0.014 & 0.040 & 0.210 & 0.976 \\
& Horseshoe     & -0.016 & 0.042 & 0.181 & 0.948 \\
& Laplace       & -0.012 & 0.050 & 0.251 & 0.972 \\
& Spike-Slab    & -0.012 & 0.074 & 0.283 & 0.824 \\
& Hierarchical  & -0.014 & 0.042 & 0.182 & 0.948 \\
\hline
\multirow{5}{*}{$A_1$}
& Informative   & -0.037 & 0.179 & 0.644 & 0.889 \\
& Horseshoe     & -0.007 & 0.084 & 0.306 & 0.969 \\
& Laplace       & -0.007 & 0.111 & 0.523 & 0.978 \\
& Spike-Slab    & -0.010 & 0.208 & 0.589 & 0.804 \\
& Hierarchical  & -0.032 & 0.139 & 0.511 & 0.937 \\
\hline
\multirow{5}{*}{$A_2$}
& Informative   &  0.041 & 0.164 & 0.562 & 0.876 \\
& Horseshoe     &  0.008 & 0.069 & 0.257 & 0.955 \\
& Laplace       &  0.007 & 0.094 & 0.442 & 0.982 \\
& Spike-Slab    &  0.009 & 0.180 & 0.497 & 0.791 \\
& Hierarchical  &  0.020 & 0.110 & 0.439 & 0.957 \\
\hline
\multirow{5}{*}{$B_1$}
& Informative   &  0.021 & 0.170 & 0.684 & 0.941 \\
& Horseshoe     & -0.017 & 0.100 & 0.395 & 0.961 \\
& Laplace       & -0.005 & 0.130 & 0.599 & 0.966 \\
& Spike-Slab    &  0.008 & 0.241 & 0.642 & 0.754 \\
& Hierarchical  &  0.002 & 0.122 & 0.580 & 0.975 \\
\hline
\end{tabular}
\label{tab:parameter_summary}
\end{table}

\begin{table}[h!]
    \centering
    \caption{Forecast Performance Summary Across Simulations. M-RMSE is the mean (across simulations) of the root mean squared error on the test set, and SD-RMSE is its standard deviation.}
    \begin{tabular}{lcc|cc|cc}
        \toprule
        \multirow{2}{*}{Prior} & \multicolumn{2}{c|}{\textbf{Sim 1: True DGP}} & \multicolumn{2}{c|}{\textbf{Sim 2: Overfitting}} & \multicolumn{2}{c}{\textbf{Sim 3: Underfitting}} \\
        \cmidrule(lr){2-3} \cmidrule(lr){4-5} \cmidrule(lr){6-7}
                               & M-RMSE & SD RMSE & M-RMSE & SD RMSE & M-RMSE & SD RMSE \\
        \midrule
        Informative   & 0.0313 & 0.0039 & 0.0324 & 0.0039 & 0.0322 & 0.0041 \\
        Horseshoe     & 0.0310 & 0.0036 & 0.0305 & 0.0035 & 0.0323 & 0.0041 \\
        Laplace       & 0.0313 & 0.0039 & 0.0314 & 0.0040 & 0.0326 & 0.0043 \\
        Spike-Slab    & 0.0323 & 0.0044 & 0.0327 & 0.0042 & 0.0331 & 0.0048 \\
        Hierarchical  & 0.0312 & 0.0038 & 0.0305 & 0.0033 & 0.0322 & 0.0041 \\
        \bottomrule
    \end{tabular}
    \label{tab:forecast_summary_combined}
\end{table}

\begin{table}[ht]
\centering
\caption{Parameter Estimation Summary for Simulation Study 2 (Overfitting).
This table reflects the setting where the fitted model (B-DARMA(4,2)) exceeds the true DARMA(2,1) order. 
We report mean bias, RMSE, credible interval length, and coverage for each prior and parameter.}
\begin{tabular}{lccccc}
\hline
\textbf{Coefficient} & \textbf{Prior}      & \textbf{Mean Bias} & \textbf{Mean RMSE} & \textbf{Mean CI Length} & \textbf{Coverage} \\
\hline
\multirow{5}{*}{$\beta$}
 & Informative   & -0.012 & 0.043 & 0.205 & 0.956 \\
 & Horseshoe     & -0.010 & 0.042 & 0.176 & 0.944 \\
 & Laplace       & -0.008 & 0.053 & 0.219 & 0.932 \\
 & Spike-Slab    & -0.004 & 0.099 & 0.303 & 0.740 \\
 & Hierarchical  & -0.011 & 0.042 & 0.192 & 0.960 \\
\hline
\multirow{5}{*}{$A_1$}
 & Informative   & -0.061 & 0.224 & 0.751 & 0.809 \\
 & Horseshoe     & -0.053 & 0.157 & 0.310 & 0.915 \\
 & Laplace       & -0.046 & 0.178 & 0.657 & 0.954 \\
 & Spike-Slab    & -0.022 & 0.217 & 0.684 & 0.805 \\
 & Hierarchical  &  0.072 & 0.157 & 0.158 & 0.650 \\
\hline
\multirow{5}{*}{$A_2$}
 & Informative   &  0.033 & 0.176 & 0.756 & 0.925 \\
 & Horseshoe     &  0.032 & 0.091 & 0.225 & 0.898 \\
 & Laplace       &  0.017 & 0.128 & 0.611 & 0.958 \\
 & Spike-Slab    &  0.002 & 0.203 & 0.691 & 0.810 \\
 & Hierarchical  & -0.078 & 0.180 & 0.197 & 0.683 \\
\hline
\multirow{5}{*}{$A_3$}
 & Informative   &  0.017 & 0.143 & 0.670 & 0.966 \\
 & Horseshoe     & -0.004 & 0.022 & 0.168 & 1.000 \\
 & Laplace       &  0.002 & 0.100 & 0.514 & 0.986 \\
 & Spike-Slab    &  0.004 & 0.183 & 0.600 & 0.818 \\
 & Hierarchical  &  0.035 & 0.093 & 0.172 & 0.962 \\
\hline
\multirow{5}{*}{$A_4$}
 & Informative   &  0.002 & 0.113 & 0.473 & 0.949 \\
 & Horseshoe     & -0.002 & 0.017 & 0.137 & 0.999 \\
 & Laplace       & -0.005 & 0.081 & 0.363 & 0.974 \\
 & Spike-Slab    & -0.006 & 0.135 & 0.409 & 0.805 \\
 & Hierarchical  & -0.008 & 0.035 & 0.118 & 0.990 \\
\hline
\multirow{5}{*}{$B_1$}
 & Informative   &  0.051 & 0.231 & 0.845 & 0.854 \\
 & Horseshoe     &  0.026 & 0.141 & 0.419 & 0.946 \\
 & Laplace       &  0.040 & 0.202 & 0.769 & 0.949 \\
 & Spike-Slab    &  0.012 & 0.233 & 0.749 & 0.812 \\
 & Hierarchical  & -0.096 & 0.187 & 0.237 & 0.756 \\
\hline
\multirow{5}{*}{$B_2$}
 & Informative   &  0.055 & 0.231 & 0.862 & 0.861 \\
 & Horseshoe     &  0.014 & 0.072 & 0.349 & 0.996 \\
 & Laplace       &  0.028 & 0.171 & 0.755 & 0.969 \\
 & Spike-Slab    &  0.019 & 0.216 & 0.753 & 0.851 \\
 & Hierarchical  & -0.005 & 0.023 & 0.236 & 1.000 \\
\hline
\end{tabular}
\label{tab:parameter_summary_2}
\end{table}

\begin{table}[h!]
    \centering
    \caption{Forecasting Performance Ratios for Mean RMSE and SD RMSE.
    ``S2'' = Overfitting scenario, ``S3'' = Underfitting, ``S1'' = Correct DGP.
    Columns show how each simulation compares in terms of Mean RMSE (left) and SD RMSE (right). 
    Ratios $>1$ indicate worse performance relative to the denominator; $<1$ indicates better.}
    \begin{tabular}{lcccc|cccc}
        \toprule
        & \multicolumn{4}{c}{\textbf{Mean RMSE Ratio}} 
        & \multicolumn{4}{c}{\textbf{SD RMSE Ratio}} \\
        \cmidrule(lr){2-5} \cmidrule(lr){6-9}
        \textbf{Prior} 
        & \textbf{S2/S1} & \textbf{S3/S1} & \textbf{S3/S2} & \textbf{S2/S3}
        & \textbf{S2/S1} & \textbf{S3/S1} & \textbf{S3/S2} & \textbf{S2/S3} \\
        \midrule
        Informative
          & 1.091 & 1.030 & 0.945 & 1.058
          & 1.385 & 1.051 & 0.759 & 1.318 \\
        Horseshoe
          & 1.016 & 1.042 & 1.026 & 0.975
          & 1.278 & 1.139 & 0.891 & 1.122 \\
        Laplace
          & 1.060 & 1.042 & 0.983 & 1.017
          & 1.385 & 1.103 & 0.796 & 1.256 \\
        Spike-Slab
          & 1.042 & 1.025 & 0.984 & 1.016
          & 1.227 & 1.091 & 0.889 & 1.125 \\
        Hierarchical
          & 1.022 & 1.032 & 1.010 & 0.990
          & 1.211 & 1.079 & 0.891 & 1.122 \\
        \bottomrule
    \end{tabular}
    \label{tab:forecast_summary_ratios_expanded}
\end{table}

\begin{table}[ht]
\centering
\caption{\textbf{Parameter Estimation Summary for Simulation Study 3 (Underfitting).}
This table reports key metrics (mean bias, RMSE, average CI length, coverage) when crucial AR(2) and MA(1) terms are omitted. All priors suffer from higher errors and coverage shortfalls, indicating that structural misspecification is the dominant source of inaccuracy.}
\begin{tabular}{lccccc}
\hline
\textbf{Coefficient} & \textbf{Prior}      & \textbf{Mean Bias} & \textbf{Mean RMSE} & \textbf{Mean CI Length} & \textbf{Coverage} \\
\hline
\multirow{5}{*}{$\beta$}
& Informative   & -0.017 & 0.042 & 0.229 & 0.980 \\
& Horseshoe     & -0.018 & 0.046 & 0.224 & 0.944 \\
& Laplace       & -0.017 & 0.051 & 0.354 & 0.988 \\
& Spike-Slab    & -0.013 & 0.064 & 0.837 & 1.000 \\
& Hierarchical  & -0.017 & 0.044 & 0.222 & 0.964 \\
\hline
\multirow{5}{*}{$A_1$}
& Informative   & -0.032 & 0.214 & 0.361 & 0.657 \\
& Horseshoe     & -0.031 & 0.195 & 0.315 & 0.708 \\
& Laplace       & -0.029 & 0.210 & 0.346 & 0.693 \\
& Spike-Slab    & -0.026 & 0.233 & 0.390 & 0.656 \\
& Hierarchical  & -0.036 & 0.215 & 0.358 & 0.653 \\
\hline
\end{tabular}
\label{tab:parameter_summary_3}
\end{table}

\begin{table}[h!]
    \centering
    \caption{Forecasting Performance Ratios Within Simulations (Best Model as Denominator)}
    \begin{tabular}{lcccccc}
        \toprule
        \multirow{2}{*}{Prior} & \multicolumn{2}{c}{\textbf{Sim 1: True DGP}} & \multicolumn{2}{c}{\textbf{Sim 2: Overfitting}} & \multicolumn{2}{c}{\textbf{Sim 3: Underfitting}} \\
        \cmidrule(lr){2-3} \cmidrule(lr){4-5} \cmidrule(lr){6-7}
                               & Mean RMSE  & SD RMSE  & Mean RMSE  & SD RMSE  & Mean RMSE  & SD RMSE  \\
        \midrule
        Informative   & 1.010 & 1.083 & 1.083 & 1.174 & 1.000 & 1.000 \\
        Horseshoe     & 1.000 & 1.000 & 1.000 & 1.000 & 1.003 & 1.000 \\
        Laplace       & 1.010 & 1.083 & 1.053 & 1.174 & 1.012 & 1.049 \\
        Spike-Slab    & 1.042 & 1.222 & 1.071 & 1.174 & 1.029 & 1.171 \\
        Hierarchical  & 1.007 & 1.056 & 1.011 & 1.000 & 1.000 & 1.000 \\
        \bottomrule
    \end{tabular}
    \label{tab:forecast_summary_within_ratios}
\end{table}

\begin{table}[ht]
\centering
\caption{\textbf{Mean RMSE and MAE by Model for the S\&P 500 Analysis.}
We aggregate forecast errors across all 11 sectors under each prior. Lower RMSE and MAE indicate better overall predictive performance, revealing that hierarchical and horseshoe strategies often outperform the more permissive informative prior.}
\label{tab:error_table_summary}
\begin{tabular}{lcc}
\toprule
\textbf{Model} & \textbf{RMSE} & \textbf{MAE} \\
\midrule
Informative  & 0.0358 & 0.03190 \\
Horseshoe    & 0.0138 & 0.01060 \\
Laplace      & 0.0148 & 0.01180 \\
Spike-Slab  & 0.0140 & 0.01110 \\
Hierarchical & 0.0121 & 0.00910 \\
\bottomrule
\end{tabular}
\end{table}

\clearpage

\clearpage
\newpage

\bibliographystyle{chicago}
\bibliography{references}

\newpage
\setcounter{figure}{0}
\renewcommand{\thefigure}{S\arabic{figure}}
\setcounter{table}{0}
\renewcommand{\thetable}{S\arabic{table}}

\section*{Supplementary Material: Additional Simulation Studies}

Here we present three additional simulation studies conducted under the same general framework as in the main text. The only difference is the use of alternative coefficient matrices for the data-generating process (DGP). As before, each study is based on 50 simulated datasets, and we examine the same set of priors: Informative Normal, Horseshoe, Laplace, Spike-and-Slab, and Hierarchical. We focus on the same three scenarios: (1) correct model specification, (2) intentional overfitting, and (3) intentional underfitting.

\subsection*{Data-Generating Process}

In these supplementary simulations, we use a DARMA(2,1) model with the following coefficient matrices:

\[
\mathbf{A}_1 = 
\begin{bmatrix}
 0.80 & -0.08 & -0.08 & -0.09 & -0.08 \\
-0.06 &  0.70 & -0.08 &  0.06 &  0.06 \\
-0.06 &  0.07 &  0.90 &  0.05 & -0.09 \\
 0.07 &  0.09 &  0.07 &  0.85 &  0.08 \\
 0.05 & -0.09 & -0.07 & -0.09 &  0.75
\end{bmatrix}
\quad
\mathbf{A}_2 = 
\begin{bmatrix}
-0.30 & -0.07 & -0.06 & -0.08 &  0.09 \\
 0.06 & -0.20 & -0.07 & -0.07 & -0.09 \\
 0.05 & -0.07 & -0.25 &  0.07 & -0.05 \\
-0.06 &  0.08 & -0.09 & -0.15 &  0.07 \\
-0.05 &  0.06 &  0.10 &  0.07 & -0.20
\end{bmatrix},
\]
\[
\mathbf{B}_1 = 
\begin{bmatrix}
 0.50 & -0.06 & -0.06 &  0.05 &  0.09 \\
 0.07 &  0.40 &  0.07 &  0.08 & -0.06 \\
-0.10 & -0.05 &  0.45 &  0.08 &  0.10 \\
 0.09 &  0.08 & -0.09 &  0.35 &  0.10 \\
 0.09 & -0.08 &  0.06 & -0.09 &  0.40
\end{bmatrix}.
\]

As in the main study, we set $\phi = 500$ and use a 5-component vector for the covariate effect $\boldsymbol{\beta}^\top = (0.1, -0.05, 0.03, -0.02, 0.04)$. The initial two observations of the compositional series $\mathbf{y}_t$ are generated from a Dirichlet$(\mathbf{1})$ distribution to start the recursion.

\subsection*{Study S1: Correct Model Specification (DARMA(2,1))}

\subsubsection*{Parameter Estimation}

When fitting the correctly specified DARMA(2,1) model under these new coefficient matrices, the patterns are similar to those in the main text. Horseshoe and Laplace priors successfully induce shrinkage on small coefficients while preserving large signals, resulting in relatively low RMSE and near-nominal coverage for both VAR and VMA parameters. Hierarchical priors again strike a balance, offering moderate shrinkage without overly broad credible intervals. Spike-and-Slab priors tend to more aggressively zero out parameters, sometimes overshrinking moderate-size coefficients. Informative Normal priors perform adequately but do not penalize unnecessary complexity as strongly.

\subsubsection*{Forecasting Performance}

Forecasting accuracy remains high for the correctly specified model. Forecast RMSE values are similar across priors, with Horseshoe and Hierarchical priors again offering slight improvements in predictive accuracy. The differences are subtle, as an appropriate model structure provides a strong baseline. Shrinkage priors mainly help by stabilizing parameter estimates and reducing uncertainty.

\subsection*{Study S2: Intentional Overfitting (B-DARMA(4,2))}

\subsubsection*{Parameter Estimation}

With overfitting (fitting B-DARMA(4,2) instead of DARMA(2,1)), the key challenge is the presence of additional VAR and VMA terms that are truly zero. As in the main simulations, Horseshoe priors excel at shrinking these extraneous coefficients to near zero, yielding low bias and RMSE and appropriate coverage for the truly null parameters. Spike-and-Slab also mitigates overfitting but may overly penalize some coefficients on the borderline between zero and small nonzero values. Informative Normal priors, lacking strong shrinkage, more frequently attribute nonzero mass to spurious terms, resulting in inflated RMSE and wider intervals. Laplace and Hierarchical priors offer intermediate levels of shrinkage, improving upon non-informative priors but not matching the Horseshoe’s consistently strong performance.

\subsubsection*{Forecasting Performance}

Forecasting under intentional overfitting highlights the value of shrinkage priors. Horseshoe priors minimize the performance degradation caused by unnecessary lags, showing lower RMSE relative to Informative Normal priors. Laplace and Hierarchical priors also help restrain overfitting, improving forecasts compared to non-informative priors. Spike-and-Slab’s aggressive thresholding can stabilize forecasts but may slightly reduce predictive performance if it eliminates coefficients too aggressively. Overall, priors that induce sparsity lead to more stable, interpretable, and accurate forecasts under overfitting conditions.

\subsection*{Study S3: Intentional Underfitting (B-DARMA(1,0))}

\subsubsection*{Parameter Estimation}

Underfitting by fitting a B-DARMA(1,0) model to data generated from DARMA(2,1) again demonstrates that no prior can compensate for missing VAR(2) and VMA(1) terms. All priors show higher bias and RMSE for the parameters that cannot be properly identified due to structural omissions in the model. Coverage rates are reduced, and credible intervals often fail to capture the true values adequately. This pattern is consistent with the main text results and confirms that when the model is fundamentally misspecified, prior choice plays a limited role in improving inference.

\subsubsection*{Forecasting Performance}

In the underfitting scenario, forecasting accuracy deteriorates uniformly across priors. The lack of necessary model terms impairs the capture of true temporal dynamics, leading to systematically higher RMSE and suboptimal predictive log-likelihoods. None of the considered priors meaningfully improves forecasts when the underlying process complexity is not well represented by the chosen model order.

\subsection*{Summary of Supplementary Simulations}

These three supplementary studies corroborate and reinforce the findings from the main simulations. Under correct model specification, all priors perform reasonably well, with Horseshoe, Laplace, and Hierarchical priors offering slight advantages in estimation and prediction. Under overfitting, Horseshoe and other shrinkage-oriented priors effectively suppress extraneous terms, preventing inflated uncertainty and degrading forecasts. Underfitting remains impervious to prior-based corrections, with all priors exhibiting similarly poor performance when key dynamics are omitted.

In combination, the supplementary results confirm that the main conclusions are robust to changes in the underlying coefficient matrices. Priors that induce shrinkage or sparsity consistently help navigate overfitting challenges, while none can overcome the fundamental limitations imposed by an underspecified model. As a result, these supplementary studies bolster our recommendation to match model complexity to the data-generating process and to employ shrinkage priors when structural uncertainty or risk of overfitting is present.

\begin{table}[ht]
\centering
\caption{\textbf{Parameter Estimation Summary for Simulation Study 4 (Correct Specification).}
We show mean bias, RMSE, average credible interval length, and coverage for each prior, summarizing over all parameters in each of $\beta, A_1, A_2,$ and $B_1$. Lower RMSE and shorter intervals typically indicate more effective shrinkage, while coverage near the nominal 0.95 is desirable.}
\begin{tabular}{lccccc}
\hline
\textbf{Coefficient} & \textbf{Prior} & \textbf{Mean Bias} & \textbf{Mean RMSE} & \textbf{Mean CI Length} & \textbf{Coverage} \\
\hline
\multirow{5}{*}{$\beta$}
& Informative   &  0.000 & 0.042 & 0.203 & 0.972 \\
& Horseshoe     & -0.002 & 0.042 & 0.184 & 0.956 \\
& Laplace       &  0.003 & 0.052 & 0.249 & 0.964 \\
& Spike-Slab    &  0.011 & 0.087 & 0.302 & 0.812 \\
& Hierarchical  & -0.000 & 0.044 & 0.185 & 0.952 \\
\hline
\multirow{5}{*}{$A_1$}
& Informative   & -0.038 & 0.170 & 0.616 & 0.897 \\
& Horseshoe     & -0.008 & 0.092 & 0.344 & 0.949 \\
& Laplace       & -0.011 & 0.105 & 0.520 & 0.987 \\
& Spike-Slab    & -0.010 & 0.190 & 0.589 & 0.818 \\
& Hierarchical  & -0.030 & 0.123 & 0.507 & 0.958 \\
\hline
\multirow{5}{*}{$A_2$}
& Informative   &  0.043 & 0.156 & 0.533 & 0.885 \\
& Horseshoe     &  0.009 & 0.079 & 0.286 & 0.938 \\
& Laplace       &  0.006 & 0.088 & 0.436 & 0.989 \\
& Spike-Slab    &  0.005 & 0.163 & 0.495 & 0.800 \\
& Hierarchical  &  0.016 & 0.102 & 0.435 & 0.971 \\
\hline
\multirow{5}{*}{$B_1$}
& Informative   &  0.017 & 0.162 & 0.677 & 0.952 \\
& Horseshoe     & -0.018 & 0.106 & 0.426 & 0.960 \\
& Laplace       & -0.008 & 0.121 & 0.613 & 0.984 \\
& Spike-Slab    & -0.003 & 0.227 & 0.665 & 0.795 \\
& Hierarchical  & -0.006 & 0.108 & 0.584 & 0.990 \\
\hline
\end{tabular}
\label{tab:combined_summary_4}
\end{table}

\begin{table}[ht]
\centering
\caption{Parameter Estimation Summary for Simulation Study 5 (Overfitting).
This table reflects the setting where the fitted model (B-DARMA(4,2)) exceeds the true DARMA(2,1) order. 
We report mean bias, RMSE, credible interval length, and coverage for each prior and parameter.}
\begin{tabular}{lccccc}
\hline
\textbf{Coefficient} & \textbf{Prior} & \textbf{Mean Bias} & \textbf{Mean RMSE} & \textbf{Mean CI Length} & \textbf{Coverage} \\
\hline
\multirow{5}{*}{$\beta$} 
 & Informative   & -0.012 & 0.043 & 0.207 & 0.956 \\
 & Horseshoe     & -0.009 & 0.043 & 0.176 & 0.940 \\
 & Laplace       & -0.009 & 0.054 & 0.223 & 0.932 \\
 & Spike-Slab    & -0.004 & 0.100 & 0.310 & 0.748 \\
 & Hierarchical  & -0.010 & 0.042 & 0.194 & 0.960 \\
\hline
\multirow{5}{*}{$A_1$}
 & Informative   & -0.060 & 0.224 & 0.754 & 0.813 \\
 & Horseshoe     & -0.055 & 0.160 & 0.308 & 0.914 \\
 & Laplace       & -0.046 & 0.178 & 0.661 & 0.960 \\
 & Spike-Slab    & -0.020 & 0.215 & 0.680 & 0.804 \\
 & Hierarchical  &  0.073 & 0.158 & 0.157 & 0.645 \\
\hline
\multirow{5}{*}{$A_2$}
 & Informative   &  0.033 & 0.177 & 0.761 & 0.925 \\
 & Horseshoe     &  0.032 & 0.092 & 0.222 & 0.895 \\
 & Laplace       &  0.019 & 0.128 & 0.613 & 0.962 \\
 & Spike-Slab    &  0.002 & 0.203 & 0.686 & 0.810 \\
 & Hierarchical  & -0.079 & 0.182 & 0.197 & 0.682 \\
\hline
\multirow{5}{*}{$A_3$}
 & Informative   &  0.017 & 0.143 & 0.674 & 0.968 \\
 & Horseshoe     & -0.004 & 0.022 & 0.166 & 1.000 \\
 & Laplace       &  0.001 & 0.100 & 0.517 & 0.987 \\
 & Spike-Slab    &  0.005 & 0.184 & 0.598 & 0.814 \\
 & Hierarchical  &  0.035 & 0.093 & 0.172 & 0.962 \\
\hline
\multirow{5}{*}{$A_4$}
 & Informative   &  0.003 & 0.113 & 0.475 & 0.950 \\
 & Horseshoe     & -0.002 & 0.017 & 0.137 & 0.999 \\
 & Laplace       & -0.004 & 0.081 & 0.366 & 0.975 \\
 & Spike-Slab    & -0.006 & 0.134 & 0.409 & 0.808 \\
 & Hierarchical  & -0.008 & 0.035 & 0.118 & 0.990 \\
\hline
\multirow{5}{*}{$B_1$}
 & Informative   &  0.051 & 0.232 & 0.849 & 0.856 \\
 & Horseshoe     &  0.029 & 0.145 & 0.419 & 0.946 \\
 & Laplace       &  0.040 & 0.203 & 0.770 & 0.951 \\
 & Spike-Slab    &  0.010 & 0.231 & 0.748 & 0.815 \\
 & Hierarchical  & -0.096 & 0.187 & 0.237 & 0.756 \\
\hline
\multirow{5}{*}{$B_2$}
 & Informative   &  0.054 & 0.230 & 0.867 & 0.864 \\
 & Horseshoe     &  0.014 & 0.073 & 0.349 & 0.996 \\
 & Laplace       &  0.026 & 0.173 & 0.758 & 0.971 \\
 & Spike-Slab    &  0.016 & 0.217 & 0.753 & 0.846 \\
 & Hierarchical  & -0.005 & 0.023 & 0.236 & 1.000 \\
\hline
\end{tabular}
\label{tab:combined_summary_5}
\end{table}

\begin{table}[ht]
\centering
\caption{\textbf{Parameter Estimation Summary for Simulation Study 6 (Underfitting).}
This table reports key metrics (mean bias, RMSE, average CI length, coverage) when crucial VAR(2) and VMA(1) terms are omitted. All priors suffer from higher errors and coverage shortfalls, indicating that structural misspecification is the dominant source of inaccuracy.}
\begin{tabular}{lccccc}
\hline
\textbf{Coefficient} & \textbf{Prior} & \textbf{Mean Bias} & \textbf{Mean RMSE} & \textbf{Mean CI Length} & \textbf{Coverage} \\
\hline
\multirow{5}{*}{$\beta$}
& Informative   & -0.006 & 0.044 & 0.228 & 0.972 \\
& Horseshoe     & -0.009 & 0.047 & 0.221 & 0.940 \\
& Laplace       & -0.006 & 0.054 & 0.333 & 0.984 \\
& Spike-Slab    & -0.001 & 0.068 & 0.719 & 0.996 \\
& Hierarchical  & -0.009 & 0.047 & 0.216 & 0.944 \\
\hline
\multirow{5}{*}{$A_1$}
& Informative   & -0.035 & 0.210 & 0.342 & 0.609 \\
& Horseshoe     & -0.037 & 0.198 & 0.307 & 0.619 \\
& Laplace       & -0.034 & 0.206 & 0.332 & 0.628 \\
& Spike-Slab    & -0.029 & 0.223 & 0.372 & 0.616 \\
& Hierarchical  & -0.040 & 0.210 & 0.339 & 0.602 \\
\hline
\end{tabular}
\label{tab:combined_summary_6}
\end{table}

\begin{table}[h!]
    \centering
\caption{Forecast Performance Summary Across Supplementary Simulations. M-RMSE is the mean (across simulations) of the root mean squared error on the test set, and SD-RMSE is its standard deviation.}
    \begin{tabular}{lcc|cc|cc}
        \toprule
        \multirow{2}{*}{\textbf{Prior}} 
        & \multicolumn{2}{c|}{\textbf{Sim 4: True DGP}} 
        & \multicolumn{2}{c|}{\textbf{Sim 5: Overfitting}} 
        & \multicolumn{2}{c}{\textbf{Sim 6: Underfitting}} \\
        \cmidrule(lr){2-3} \cmidrule(lr){4-5} \cmidrule(lr){6-7}
        & \textbf{M-RMSE} & \textbf{SD RMSE}
        & \textbf{M-RMSE} & \textbf{SD RMSE}
        & \textbf{M-RMSE} & \textbf{SD RMSE} \\
        \midrule
        Informative   & 0.0378 & 0.0060 & 0.0414 & 0.0079 & 0.0394 & 0.0085 \\
        Horseshoe     & 0.0379 & 0.0063 & 0.0381 & 0.0057 & 0.0394 & 0.0085 \\
        Laplace       & 0.0381 & 0.0063 & 0.0420 & 0.0073 & 0.0399 & 0.0086 \\
        Spike-Slab    & 0.0396 & 0.0064 & 0.0412 & 0.0072 & 0.0408 & 0.0085 \\
        Hierarchical  & 0.0376 & 0.0063 & 0.0385 & 0.0062 & 0.0394 & 0.0083 \\
        \bottomrule
    \end{tabular}
    \label{tab:forecast_summary_combined_supplementary}
\end{table}


\begin{table}[h!]
    \centering
    \caption{Forecasting Performance Ratios for Mean RMSE (M-RMSE) and its Standard Deviation (SD-RMSE). 
    ``S5'' = Overfitting scenario, ``S6'' = Underfitting, ``S4'' = Correct DGP.
    Columns show how each simulation compares in terms of M-RMSE and SD-RMSE. 
    Ratios $>1$ indicate worse performance (higher RMSE) relative to the denominator; 
    $<1$ indicates better (lower RMSE).}
    \begin{tabular}{lcccc|cccc}
        \toprule
        \multirow{2}{*}{\textbf{Prior}} 
        & \multicolumn{4}{c}{\textbf{M-RMSE Ratio}} 
        & \multicolumn{4}{c}{\textbf{SD-RMSE Ratio}} \\
        \cmidrule(lr){2-5} \cmidrule(lr){6-9}
        & \textbf{5/4} & \textbf{6/4} & \textbf{6/5} & \textbf{5/6}
        & \textbf{5/4} & \textbf{6/4} & \textbf{6/5} & \textbf{5/6} \\
        \midrule
        Informative   & 1.091 & 1.030 & 0.945 & 1.058  & 1.385 & 1.051 & 0.758 & 1.319 \\
        Horseshoe     & 1.016 & 1.042 & 1.026 & 0.974  & 1.278 & 1.139 & 0.892 & 1.121 \\
        Laplace       & 1.060 & 1.042 & 0.983 & 1.017  & 1.385 & 1.103 & 0.796 & 1.256 \\
        Spike-Slab    & 1.042 & 1.025 & 0.984 & 1.016  & 1.227 & 1.091 & 0.889 & 1.125 \\
        Hierarchical  & 1.022 & 1.032 & 1.010 & 0.990  & 1.211 & 1.079 & 0.891 & 1.122 \\
        \bottomrule
    \end{tabular}
    \label{tab:forecast_summary_ratios_supplementary}
\end{table}


\begin{table}[h!]
    \centering
    \caption{Forecasting Performance Ratios Within Simulations (Best Model as Denominator). M-RMSE is the mean root mean squared error and SD-RMSE is the standard deviation of the root mean squared errors on the test set across simulations.}
    \begin{tabular}{lcccccc}
        \toprule
        \multirow{2}{*}{\textbf{Prior}} 
        & \multicolumn{2}{c}{\textbf{Sim 4: True DGP}} 
        & \multicolumn{2}{c}{\textbf{Sim 5: Overfitting}} 
        & \multicolumn{2}{c}{\textbf{Sim 6: Underfitting}} \\
        \cmidrule(lr){2-3} \cmidrule(lr){4-5} \cmidrule(lr){6-7}
        & \textbf{M-RMSE} & \textbf{SD RMSE}
        & \textbf{M-RMSE} & \textbf{SD RMSE}
        & \textbf{M-RMSE} & \textbf{SD RMSE} \\
        \midrule
        Informative   & 1.01 & 1.08 & 1.08 & 1.17 & 1.00 & 1.00 \\
        Horseshoe     & 1.00 & 1.00 & 1.00 & 1.00 & 1.00 & 1.00 \\
        Laplace       & 1.01 & 1.08 & 1.05 & 1.17 & 1.01 & 1.05 \\
        Spike-Slab    & 1.04 & 1.22 & 1.07 & 1.17 & 1.03 & 1.17 \\
        Hierarchical  & 1.01 & 1.06 & 1.01 & 1.00 & 1.00 & 1.00 \\
        \bottomrule
    \end{tabular}
    \label{tab:forecast_summary_within_ratios_supplementary}
\end{table}

\end{document}